\documentclass{article}
\usepackage{spconf,amsmath,graphicx}
\usepackage{booktabs, url}

\title{Unsupervised Word Segmentation \\ Using Temporal Gradient Pseudo-Labels}
\twoauthors
  {Tzeviya Sylvia Fuchs}
	{Bar-Ilan University, Ramat-Gan, Israel}
  {Yedid Hoshen}
	{The Hebrew University of Jerusalem}

\begin{document}
%
\maketitle
\begin{abstract}
Unsupervised word segmentation in audio utterances is challenging as, in speech, there is typically no gap between words. In a preliminary experiment, we show that recent deep self-supervised features are very effective for word segmentation but require supervision for training the classification head. To extend their effectiveness to unsupervised word segmentation, we propose a pseudo-labeling strategy. Our approach relies on the observation that the temporal gradient magnitude of the embeddings (i.e. the distance between the embeddings of subsequent frames) is typically minimal far from the boundaries and higher nearer the boundaries. We use a thresholding function on the temporal gradient magnitude to define a psuedo-label for wordness. We train a linear classifier, mapping the embedding of a single frame to the pseudo-label. Finally, we use the classifier score to predict whether a frame is a word or a boundary. In an empirical investigation, our method, despite its simplicity and fast run time, is shown to significantly outperform all previous methods on two datasets. Code is available at \url{https://github.com/MLSpeech/GradSeg}. 

\end{abstract}
\section{Introduction}
\label{sec:intro}

Spoken word segmentation is often the first step in language acquisition. It aims to predict all word boundaries within an utterance. Spoken language usually does not contain gaps between words, making their segmentation challenging. This is different from text, where the boundaries are explicitly marked using space characters. Unsupervised word segmentation is an important subfield, which does not assume any prior labels on word boundaries. This is often the case in language acquisition in unknown, low-resource languages. The significance of the field has led to extensive research over the past decades e.g., \cite{rasanen2009improved,rasanen2012computational,rasanen2020unsupervised,kamper2017embedded,kamper2017segmental,kamper2020towards,bhati2021segmental,johnson2009improving,cuervo2021contrastive, iwamoto2021unsupervised}. 

 Probabilistic modeling is the most popular approach for this task \cite{kamper2020towards, bhati2021segmental, johnson2009improving, cuervo2021contrastive}. Such models rely on the assumption that samples lying across different words are less correlated than two samples within the same word. This criterion is used for estimating the word boundaries. Current techniques include one- and two-stage approaches. Two-stage approaches are the most common \cite{kamper2020towards, bhati2021segmental, cuervo2021contrastive}. They first segment the utterance into its phonemes and then learn a discrete language model over the approximate phonemes. Fuchs et al. \cite{fuchs22_interspeech} recently introduced a single stage probabilistic model which does not require prior phoneme segmentation. The method first embeds the utterance into semantic representations using a pretrained self-supervised model e.g., Wav2Vec 2.0 \cite{baevski2020wav2vec}. The method uses a simple kNN estimator for the probability of the input utterance. High kNN distances were shown to be predictive of word boundaries. The top single- and two-stage methods achieve roughly similar performance. While most current approaches follow the language modeling paradigm, its assumptions are often violated as some word combinations e.g., ``they are'', often co-occur more frequently than most single words. 

In this work, we break from the standard probabilistic modeling paradigm. Instead, we hypothesize that deep pretrained self-supervised representations e.g., Wav2Vec 2.0, already contain the information needed to locate word boundaries. The hypothesis is confirmed in our preliminary experiments. The challenge is to extract this information \textit{without} having any labelled examples of word boundaries. To tackle this challenge, we start by computing the gradient between subsequent frames in the dataset. We observe an interesting empirical result; the gradient value is low within a given word, and tends to be high between different words. While this metric by itself does not outperform the state-of-the-art in unsupervised word segmentation, it can be used for pseudo-labeling. Specifically, binary pseudo-labels are created by thresholding the gradient magnitude with a constant value. The pseudo-labels are then used for training a linear classifier on top of the pretrained features. We find that the classifier score is a good predictor of word boundaries. We evaluate our method against the current state-of-the-art one- and two-stage methods on two datasets, YOHO \cite{campbell1995testing} and Buckeye \cite{pitt2005buckeye}. Our method achieves significantly better results than previous methods, while being faster and simpler.

\section{Method}
\label{sec:method}

\begin{figure*}[]
\centering
\includegraphics[width=8.5cm]{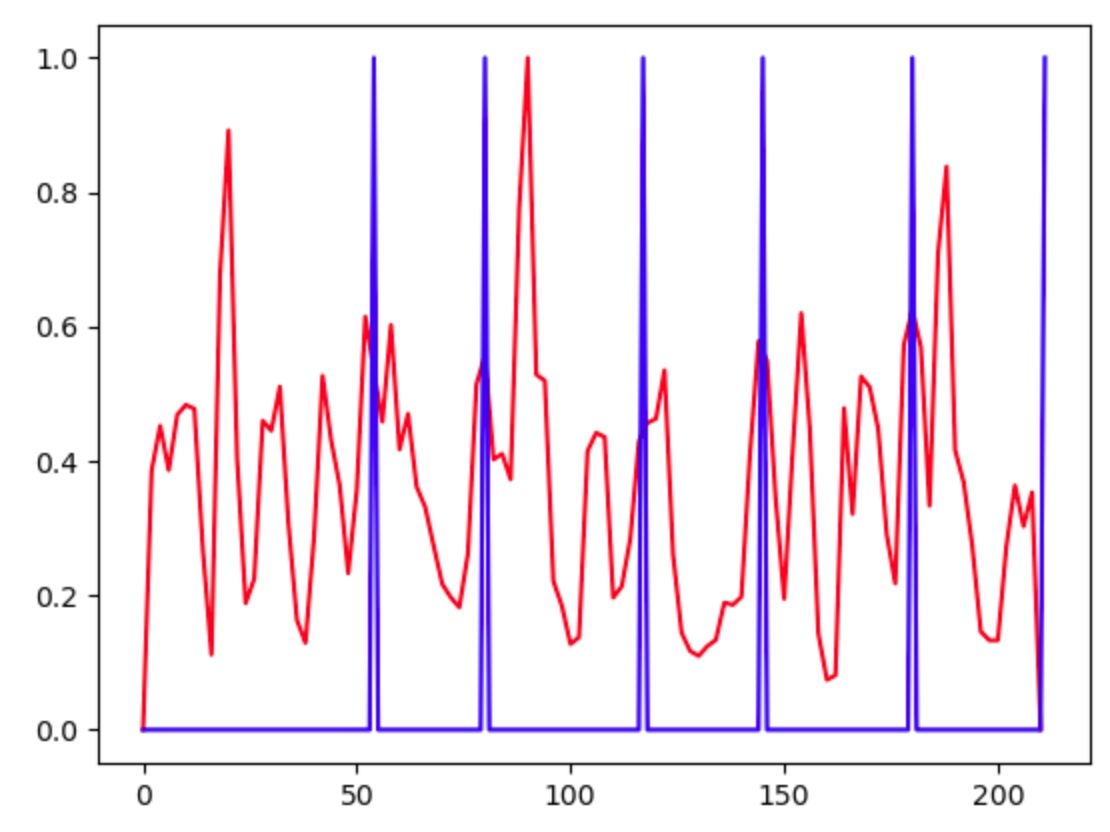}~~~~~
\includegraphics[width=8.5cm]{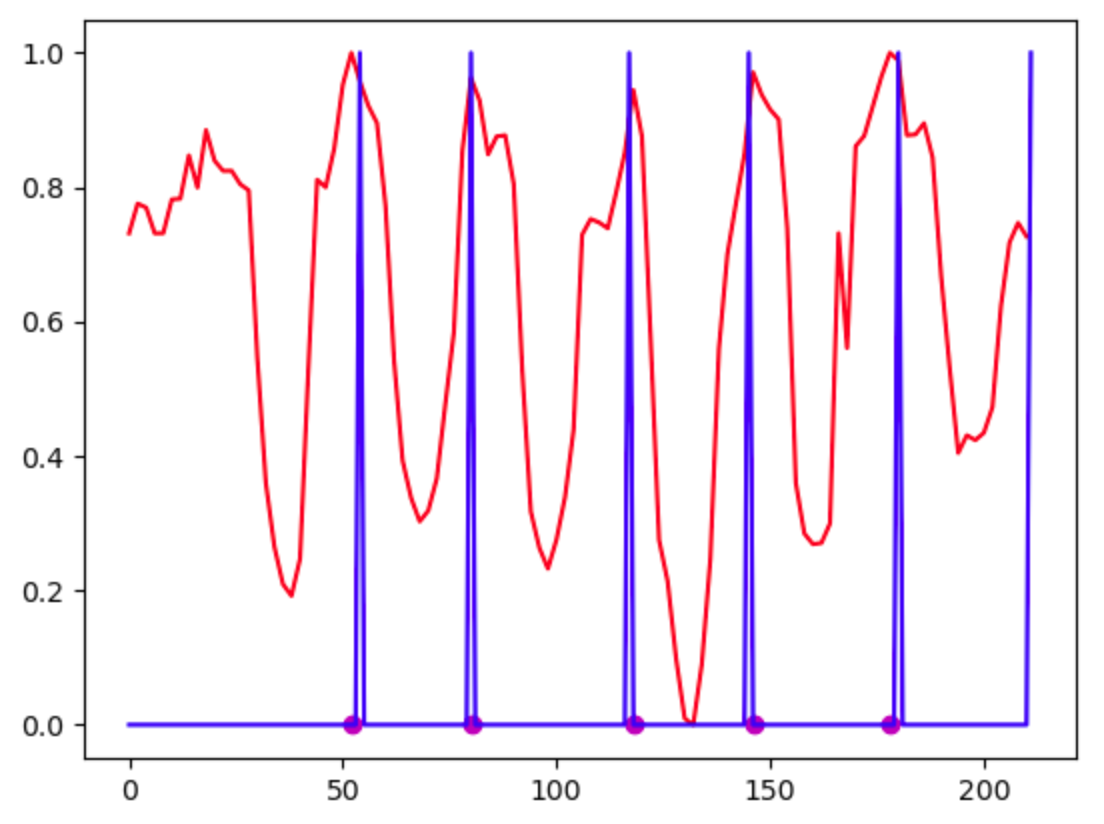}
\caption{An illustration of our method. (Left) Temporal gradient magnitude (y axis) vs. time (x axis). The blue bars indicate the groundtruth word boundaries. While the peaks do not always align with the word boundaries, the valleys are consistently far from the boundaries. This allows us to use them as pseudo labels. (Right) Classifier score (y axis) vs. time (x axis). The magenta dots are the predictions of our NMS peak detector. The predictions are roughly aligned with the word boundaries.}
\label{fig:res}
\end{figure*}

\subsection{Notation}

Let us denote the speech signal as $\bar{\mathbf{x}} = [x_1,x_2, \ldots, x_T]$, where $x_t$ denotes the sample at time $t$. The input signal $\bar{\mathbf{x}}$ is represented using a feature extractor $\phi$. The signal $\bar{\mathbf{x}}$ is thus encoded into a feature vector sequence $\bar{\mathbf{f}} = [\mathbf{f}_1,\mathbf{f}_2 \ldots, \mathbf{f}_N]$ such that $\bar{\mathbf{f}} = \phi(\bar{\mathbf{x}})$. The segmentation of the utterance is parameterized by the boundaries $\bar{\mathbf{b}} = [1,b_1, \ldots, b_M,T]$. The $m$th word starts and ends at times $b_{m-1}$ and $b_{m}$ respectively, consisting of the samples $\hat{\mathbf{x}}^m = [x_{b_{m-1}},x_{b_{m-1}+1}, \ldots,x_{b_{m}}]$.

\subsection{Self-Supervised Features for Word Segmentation}
\label{subsec:sup}

Deep self-supervised features, e.g., Wav2Vec 2.0 \cite{baevski2020wav2vec},  have recently been used to achieve state-of-the-art unsupervised word segmentation results. However, as the overall accuracy of current unsupervised word segmentation is still not high, this does not provide an understanding of the expressivity of the features. In this section, we conduct a preliminary experiment to verify this accuracy, by upper-bounding it using full supervision. Note that supervision is used in this experiment only, all other experiments in this paper \textit{do not} use training supervision. In this experiment,  we first pass the utterance through a Wav2Vec 2.0 feature extractor, resulting in deep features for every $20$ ms interval (we denote this as a ``frame''). We label each frame in the Buckeye training set as ``boundary'' or not, using the ground truth. Finally, we train a linear classifier (using ridge regression \cite{hoerl1970ridge}) mapping the deep feature to the ground truth labels. During training, we keep the feature extractor frozen without finetuning. We evaluate the accuracy of our method on the validation set using the F1-score. We obtained a value of 59.6\%, about double the state-of-the-art unsupervised word segmentation accuracy.

While it is unsurprising that supervised methods reach better performance than unsupervised ones, the results demonstrate that deep self-supervised features of even a single frame are sufficient for boundary detection. Furthermore, this information can be obtained using a simple linear classifier. The task that we will tackle in the following sections is finding a sufficiently good linear classifier without supervision.

\subsection{Connection Between Temporal Gradients and Words}
\label{subsec:grad}

In this section, we seek an unsupervised indicator for the positions of frames with respect to word boundaries. Our hypothesis is based on Sec. \ref{subsec:sup}, where we showed that powerful deep self-supervised feature extractors such as Wav2Vec 2.0 already contain this information. We analyse the temporal gradient of the embeddings which we define as the vector difference between two adjacent embeddings.
\begin{equation}
\Delta \mathbf{f}_t = \frac{\mathbf{f}_{t+1} - \mathbf{f}_{t-1}}{2}
\end{equation}

We describe each frame by the magnitude of the temporal gradient, denoted by the scalar $m_t$. It encompasses the amount of variation between the frames.

\begin{equation}
m_t = \|\Delta \mathbf{f}_t\|^2
\end{equation}

\textbf{Finding 1: Temporal gradient magnitude is a mediocre word segmenter.} To empirically evaluate the gradient magnitude as a word segmenter, we obtain a set of word boundaries from the gradient vector.  We implement a simple peak detector that searches for the highest peaks, while also ensuring non-maxima suppression (see Sec.~\ref{subsec:pseudo} for precise details). The output of our full method is a set of predicted boundaries $\bar{\mathbf{b}}$ for each input utterance $\bar{\mathbf{x}}$. In our evaluation, it achieved F1-scores of $13.2 \%$ (YOHO) and $21.3 \%$ (Buckeye). These results do not compare favorably with the state-of-the-art, but are comparable to some older previous works (see Tab.~\ref{tab:buckeye_compare}). 

\textbf{Finding 2: Low temporal gradient magnitudes are excellent at detecting far-from-boundary regions.} We visualize the temporal gradient magnitude for a single utterance in Fig.~\ref{fig:res}(left). We observe that while high temporal magnitudes do not always align with boundary regions, low values of gradient magnitudes correlate with regions that are far from the boundary. This is a useful observation that will be exploited in the next section.

\subsection{Our Approach: Pseudo-Labeling}
\label{subsec:pseudo}

We propose a new approach, GradSeg, for extracting word boundaries without supervision. Our approach utilizes the observation we made in the previous section, that very low gradient magnitudes are predictive of far-from-boundary frames. Concretely, we utilize the gradient magnitudes $m_t$ to create pseudo-labels for the data. Specifically, we threshold each gradient magnitude $m_t$ with threshold $\theta$. Frames with magnitudes smaller than $\theta$ are given positive labels (``far-from-boundary word''), while those with larger values have negative labels (unknown).
\begin{equation}
p_t = 1_{[m_t > \theta]}
\end{equation}

The detection threshold is set to the lowest $20$th percentile of the gradient values in the training set. We train a linear classifier on the frozen pretrained features $\mathbf{f}_t$, mapping them to the pseudo labels $p_t$. After training the classifier $c$, we obtain a prediction score $s_t = c(\mathbf{f}_t)$ for time $t$.  

\textbf{Non-maxima suppression.} We extract the peak values using a non-maxima-suppression (NMS) type method \cite{neubeck2006efficient}. We first rank all the frames of a particular utterance using their scores $s_t$.  We define a set of selected boundaries, which is initially empty. We compute the desired number of words by the duration of the utterance, divided by the duration $\tau_{avg}$ of an average word in the target language.  We then iterate for each frame, from the highest to the lowest score. If the minimal distance between the frame to all selected boundaries is larger than the minimal word duration $\tau_{min}$ and the desired number of words has not been exceeded, then it is added to the list of selected boundaries. Otherwise, it is skipped. As an illustration, we present the results of our full method on an example utterance in Fig.~\ref{fig:res}~(right).

\section{Experiments}
\label{sec:exp}

We use two datasets to evaluate our method, the Buckeye corpus \cite{pitt2005buckeye} and the YOHO speaker verification dataset \cite{campbell1995testing}. 

\textbf{Buckeye \cite{pitt2005buckeye}.} We follow the setup in \cite{kreuk2020self}. The Buckeye corpus consists of conversational speech of $40$ speakers. We use a train/val/test split of $0.8$/$0.1$/$0.1$. Long speech sequences are divided into short segments by splitting at noisy or silent intervals. We keep 20ms of silence before and after each segment. Similarly to the protocol of previous work \cite{kamper2020towards, bhati2021segmental, fuchs22_interspeech}, the development set (7 hours) is used for evaluation.

\textbf{YOHO \cite{campbell1995testing}.} Each utterance of the YOHO dataset contains a sequence of three numbers, each consisting of two digits. The digits are uttered by 138 speakers. A split of  $0.8$/$0.1$/$0.1$ is performed for the train/val/test sets respectively. Each speaker is assigned to one split only. While precisely aligned transcripts are not provided by the dataset, we use the Montreal Forced Aligner (MFA) \cite{mcauliffe2017montreal} to provide such alignment. We removed the silence at the beginning and end of each utterance using a voice activity detector (VAD)\footnote{{\texttt{https://github.com/wiseman/py-webrtcvad}}}. The processed validation data consisted of around 1 hour. 

\textbf{Metrics.} We report precision, recall, F1-score, over-segmentation (OS) and R-value \cite{rasanen2009improved}. These metrics are commonly used by unsupervised word segmentation papers including \cite{kamper2020towards, fuchs22_interspeech}. OS evaluates whether more or fewer boundaries are predicted than the groundtruth. The desired value of OS is $0$ as both large negative and positive values indicate imperfect results. The R-value weighs recall and OS. A value of $1$ is obtained when both metrics have perfect scores.

\subsection{Evaluation}

\begin{table}
\renewcommand{\arraystretch}{1}
  \centering
  \resizebox{\columnwidth}{!}{
  \begin{tabular}{l c c c c c}
    \hline
   \multicolumn{1}{c}{Model} & \multicolumn{1}{c}{\textbf{Prec.}} &  \multicolumn{1}{c}{\textbf{Recall}} &  \multicolumn{1}{c}{\textbf{F-score}} &  \multicolumn{1}{c}{\textbf{OS}}  &  \multicolumn{1}{c}{\textbf{R-val}}   \\
    \hline
    ES-KMeans \cite{kamper2017embedded} &  30.7    &  18.0   & 22.7  & -41.2 & 39.7\\
    BES-GMM \cite{kamper2017segmental} &   31.7   &   13.8  &  19.2 & -56.6 & 37.9\\
    VQ-CPC DP \cite{kamper2020towards} &   15.5   &  \bf{81.0}   & 26.1 & 421.4  & -266.6\\
    VQ-VAE DP \cite{kamper2020towards} &     15.8  & 68.1 & 25.7  &  330.9  & -194.5 \\
    AG VQ-CPC DP \cite{kamper2020towards} &   18.2   &  54.1   &  27.3 & 196.4 & -86.5\\
    AG VQ-VAE DP \cite{kamper2020towards} &  16.4     &    56.8  & 25.5  &  245.2  & -126.5  \\
    Buckeye\_SCPC \cite{bhati2021segmental} &   35.0   &  29.6  &  32.1  & -15.4  & 44.5 \\
    DSegKNN \cite{fuchs22_interspeech} &   30.9   &  32.0   & 31.5 & 3.46 & 40.7\\
    GradSeg  &   \bf{44.5}   &  43.6   & \bf{44.1} & \bf{-2.0} & \bf{52.6}\\
    \hline
  \end{tabular}}
 \caption{ \label{tab:buckeye_compare} Word segmentation accuracy state-of-the-art comparison (Buckeye validation F1-score - \%). GradSeg outperforms all methods by a large margin. }
\end{table}

\begin{table}
\renewcommand{\arraystretch}{1}
  \centering
  \resizebox{\columnwidth}{!}{
  \begin{tabular}{l c c c c c}
    \hline
   \multicolumn{1}{c}{Model} & \multicolumn{1}{c}{\textbf{Prec.}} &  \multicolumn{1}{c}{\textbf{Recall}} &  \multicolumn{1}{c}{\textbf{F-score}} &  \multicolumn{1}{c}{\textbf{OS}}  &  \multicolumn{1}{c}{\textbf{R-val}}   \\
    \hline
    DSegKNN \cite{fuchs22_interspeech}  &   40.8   &  \bf{45.1}   & 42.9 & 10.38 & 49.0\\
    GradSeg  &   \bf{43.8}   &    43.8 & \bf{43.8} & \bf{0.0} & \bf{51.9}\\
    \hline
  \end{tabular}}
 \caption{ \label{tab:yoho_compare} A comparison of word segmentation accuracy (\%) on the YOHO validation set. Our method GradSeg outperforms DSegKNN \cite{fuchs22_interspeech} which reports results on this dataset. }
\end{table}

We compare our method, GradSeg, to classic and state-of-the-art  methods on the Buckeye validation set (Tab.~\ref{tab:buckeye_compare}). The baseline numbers are copied from Bhati et al. \cite{bhati2021segmental}. Our method  significantly outperforms other methods in all metrics except for the recall, where the vector-quantized (VQ) based approaches achieved better results. We also compared our method to DSegKNN \cite{fuchs22_interspeech} which also reported results on the YOHO dataset. While DSegKNN performs well on the YOHO dataset (due to its very limited vocabulary size), we still outperform it on all metrics except recall.

\subsection{Hyperparameters}. 

 \noindent \textit{Architecture:} Wav2Vec2.0-Base, pre-trained on unlabeled LibriSpeech. \textit{Training data.} We used only $100$ randomly selected training utterances. This led to fast run time while retaining high accuracy. \textit{Regularization:} The ridge regression regularization parameter was $10^9$ for YOHO and $10^7$ for Buckeye. \textit{Pseudo-labelling threshold ($\theta$):} The value of $\theta$ was selected so that the lowest $20\%$ were labeled as non boundary instances. \textit{NMS peak detection:} we set the average word duration $\tau_{avg}$ to be $300$ms for Buckeye. As YOHO utterances contain $6$ words, we used this value in the NMS. For both datasets we used a minimal word duration $\tau_{min}$ of $60$ ms.

\subsection{Ablation}

\begin{figure}[t]
\centering
\includegraphics[width=6.8cm]{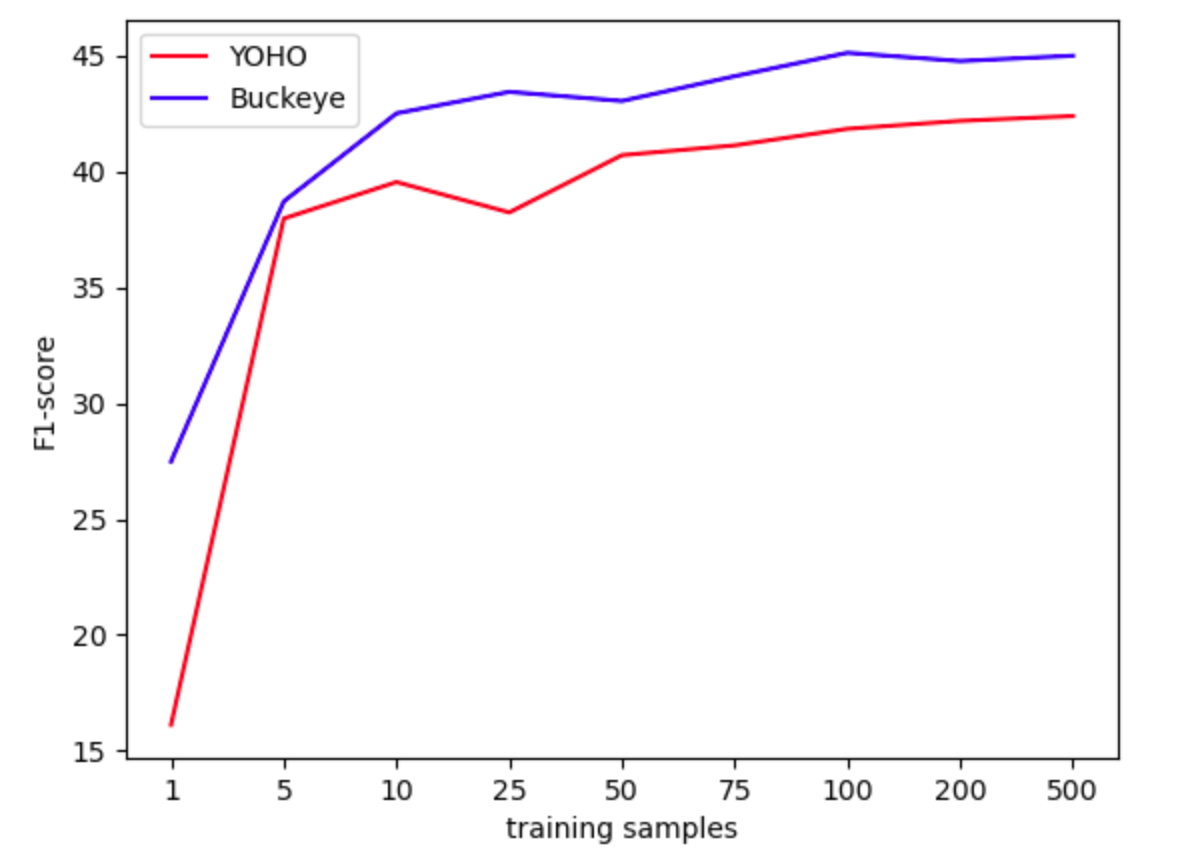} \\
\caption{Word segmentation (F1-score) as a function of the number of training samples for YOHO and Buckeye. Both datasets experience improved performance with increasing size but the gains saturate after about 100 utterances.}
\label{fig:train_n}
\end{figure}

\noindent\textbf{Training set size.} We plot the accuracy vs. number of training \textit{utterances} in Fig.~\ref{fig:train_n}. Even a small number of speech utterances is enough to train the classifier. Around $100$ utterances enjoyed the best accuracy vs. training-time tradeoff. 

\begin{figure}
\centering
\includegraphics[width=6.8cm]{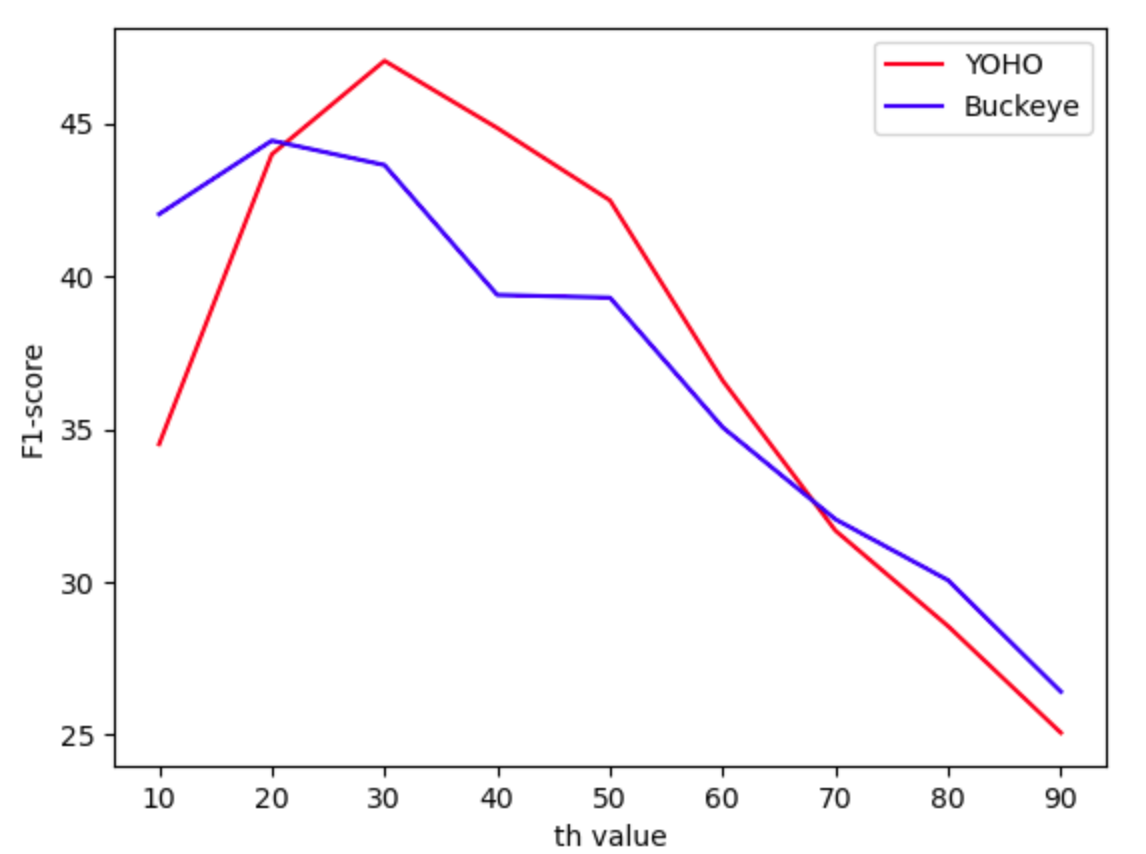} \\
\caption{Word segmentation (F1-score) vs. pseudo-label threshold  ($\theta$). Best performance is with $\theta=20\%-30\%$.}
\label{fig:threshold}
\end{figure}

\noindent\textbf{Pseudo-Label Threshold ($\theta$).} We investigated the effect of different choices of the threshold $\theta$ in Fig.~\ref{fig:threshold}. Our method achieved the best results when the lowest $20\%-30\%$ frames were labelled as negative. Results began to decline when the threshold was increase to $40\%$ or higher. A possible explanation for the lack of symmetry can be motivated by Fig.~\ref{fig:res}(left); very low gradient magnitudes are predictive of far-from-boundary frames, but high gradient magnitudes do not always indicate word boundaries. 

\noindent\textbf{Optimization objective}. We compare training with logistic vs. ridge regression. Logistic regression achieved comparable or slightly better performance than ridge regression (F1-score: 45.0 on YOHO and 44.6 on Buckeye) but has a longer training time and an extra optimization hyperparameter. We expect that exploration of classification objectives may result in further gains. As this is orthogonal to our main ideas, we left a more extensive exploration to future work. 

\noindent\textbf{Peak detector.} We introduced a new non-minima suppression based detector. This was necessary as the peaks generated by GradSeg are sometimes quite flat leading to multiple detections that need to be suppressed. To reject the hypothesis that our performance gains over DSegKNN \cite{fuchs22_interspeech} stem from the peak detector rather than the DeepSeg method, we tested DSegKNN with our new NMS peak detector. The results were not significantly different from those reported in the original paper. This is unsurprising, as DSegKNN yields rather pointy peaks making their discovery insensitive to the precise choice of peak detector.   

\subsection{Runtime Complexity Analysis }

Our method has low runtime complexity. Training requires a single evaluation of the pretrained Wav2Vec2.0 for every sample, similarly to DSegKNN \cite{fuchs22_interspeech} and unlike \cite{kamper2020towards} that requires multiple training epochs. At inference time, our method requires only a single evaluation of the feature encoder, similarly to \cite{kamper2020towards}, but unlike DSegKNN which also requires a nearest neighbor search which is expensive for large training sets. The cost of all other stages of our method is negligible.

\section{Conclusions}
\label{sec:conclusions}

We first showed that modern self-supervised features e.g., Wav2Vec2.0 can predict word boundaries even with a linear classifier but require supervision to train it. We then showed that the temporal gradient magnitude of the embeddings is a good indicator of the non-boundary region, although it is not a good word segmenter on its own. Finally, we trained a high-accuracy boundary classifier using low-gradient magnitudes as a pseudo-label. In our experiments, our method significantly outperformed the current state-of-the-art while being simpler than many previous methods.

\bibliographystyle{IEEEbib}
\bibliography{refs}

\end{document}